\documentclass{aa}
\usepackage{graphicx}
\begin{document}

\hyphenation{Fe-bru-ary Gra-na-da mo-le-cu-le mo-le-cu-les}

\title{Detection of doubly-deuterated methanol in the solar-type
protostar IRAS16293$-$2422}

\author{
   B.Parise\inst{1}
   \and C.Ceccarelli\inst{2,3}
   \and A.G.G.M.Tielens\inst{4}
   \and E.Herbst\inst{5}
   \and B.Lefloch\inst{3}
   \and E.Caux\inst{1}
   \and A.Castets\inst{2}
   \and I.Mukhopadhyay\inst{6}
   \and L.Pagani\inst{7}
   \and L.Loinard\inst{8}
}
\institute{
CESR CNRS-UPS, BP 4346, 31028 - Toulouse cedex 04, France
\and
Observatoire de Bordeaux, BP 89, 33270 Floirac, France
\and
Laboratoire d'Astrophysique, Observatoire de Grenoble -BP 53, F-38041
Grenoble cedex 09, France
\and
SRON, P.O. Box 800, NL-9700 AV Groningen, the Netherlands
\and
Department of Physics, The Ohio State University, 174 W. 18th Ave.
Columbus, OH 43210-1106, USA
\and
College of Natural Sciences, Dakota State University, Madison, SD 57042-1799, USA
\and
LERMA \& FRE 2460 du CNRS, Observatoire de Paris, 61 Av. de
l'Observatoire de Paris, 75014 Paris, France
\and
Instituto de Astronom\'{\i}a, UNAM,
   Apdo Postal 72-3 (Xangari), 58089 Morelia, Michoac\'an, M\'exico}
\offprints{Berengere.Parise@cesr.fr}

\date{Received {\today} /Accepted }
\titlerunning{Deuterated methanol in IRAS16293-2422}
\authorrunning{Parise et al.}

\abstract{We report the first detection of doubly-deuterated methanol
(CHD$_2$OH), as well as firm detections of the two singly-deuterated
isotopomers of methanol (CH$_2$DOH and CH$_3$OD), towards the
solar-type protostar IRAS16293$-$2422.  From the present multifrequency
observations, we derive the following abundance ratios:
[CHD$_2$OH]/[CH$_3$OH] = $0.2 \pm 0.1$, [CH$_2$DOH]/[CH$_3$OH] = $0.9
\pm 0.3$, [CH$_3$OD]/[CH$_3$OH] = $0.04 \pm 0.02$.  The total
abundance of the deuterated forms of methanol is greater than that of
its normal hydrogenated counterpart in the circumstellar material of
IRAS16293$-$2422, a circumstance not previously encountered.
Formaldehyde, which is thought to be the chemical precursor of methanol,
possesses a much lower fraction of deuterated isotopomers ($\sim
20$\%) with respect to
the main isotopic form in IRAS16293$-$2422. The observed fractionation
of methanol and formaldehyde provides a severe challenge to both
gas-phase and grain-surface models of deuteration.  Two examples of 
the latter model
are roughly in agreement with our observations of CHD$_2$OH and CH$_2$DOH if 
the accreting gas has a large (0.2-0.3) atomic D/H ratio.  However, no
gas-phase model predicts such a high atomic D/H ratio, and hence some
key ingredient seems to be missing.
 }

\maketitle

\section{Introduction}

In the last few years, the study of doubly-deuterated molecules in
the interstellar medium has gained considerable attention.
This field was boosted by the
discovery of an extremely large amount (D$_2$CO/H$_2$CO $\sim$ 10\%)
of doubly-deuterated formaldehyde in the low mass protostar
IRAS16293$-$2422 (hereafter IRAS16293 ; \cite{Ceccarelli98} 1998), a
fractionation about 25 times larger than in Orion (\cite{Turner90} 1990).
Follow-up observational studies of this first discovery confirmed this
very large degree of deuteration in IRAS16293 (\cite{Loinard00} 2000), and
allowed a study of its spatial distribution (\cite{Ceccarelli01} 2001).
Subsequently, similarly large amounts of doubly-deuterated
formaldehyde and ammonia have been observed towards another very young
protostellar core, 16293E, which lies in the same molecular cloud
(L1689N) as IRAS16293 (\cite{Loinard01} 2001) and in the molecular cloud
L1689N itself (\cite{Ceccarelli02} 2002).  Finally, preliminary results of
an ongoing project show that it is likely that {\it all} low-mass
protostars present similarly large abundance ratios of doubly
deuterated formaldehyde with respect to H$_2$CO, whereas high-mass
protostars do not (\cite{Loinard02a} 2002, \cite{Ceccarelli02} 2002).

All these observations suggest that such a large deuteration of
formaldehyde is produced during the cold and dense pre-collapse phase
of low-mass protostars.  Highly deuterated ices are very likely formed
via active grain chemistry (\cite{Tielens83} 1983), stored on the grain
mantles, and eventually released into the gas during the collapse
phase, when the heating of newly-formed protostars evaporates the
CO-rich ices (\cite{Ceccarelli01} 2001a). Methanol is also commonly
believed to be formed on grain surfaces, because gas-phase models
cannot account for the large detected abundances of methanol in hot
cores (\cite{Menten88} 1988).  If formaldehyde and methanol are produced on
grain surfaces by simple successive hydrogenations of CO, then the
reproduction of the abundance ratios between deuterated isotopomers
and their normal counterparts is a crucial test for the grain-surface
theory of deuteration (e.g. \cite{Charnley97} 1997).

Although the grain picture seems qualitatively consistent with all the
observations so far available towards protostars, the nature of the 
production of
deuterated molecular species,  whether it occurs completely via 
active grain chemistry
(starting from a high D/H atomic ratio derived from gas-phase
chemistry)
or at least partially via gas-phase formation, is still largely
debated (see for example \cite{Roberts00b} 2000b).  The debate has not
been settled conclusively because of the relatively small body of
available observations and the discovery of relatively large
abundances of doubly-deuterated ammonia (NHD$_2$/NH$_3 \sim$
0.001; \cite{Roueff00} 2000) in the molecular cloud L134N and triply-deuterated
ammonia in the low-mass protostar NGC1333-IRAS4 (\cite{vanderTak02} 2002)
and in the dark cloud B1 (\cite{Lis02} 2002).  The observed fractionation of
ammonia in L134N can be accounted for by gas-phase models if a high
degree of depletion of heavy materials onto the grain mantles is
assumed (\cite{Roberts00b} 2000b, \cite{Rodgers01} 2001).  The ND$_{3}$ observations from
\cite{Lis02} (2002) can also be explained in the framework of gas-phase
chemical models if the dissociative recombination of partially
deuterated ions results in a somewhat higher probability for the
ejection of hydrogen atoms than for deuterium atoms.  We wish to
emphasize that this debate is not merely academic, as it involves our
understanding of the chemistry of the interstellar medium and of ice
formation in general and deuteration processes in particular.  Many
observational studies use deuteration processes, which are supposedly
well-understood, to derive key quantities such as the deuterium
abundance (e.g.  in the Galactic Center ; \cite{Lubowich00} 2000) or the
degree of ionization (e.g. in protostars ; \cite{Williams98} 1998).  The
actual state of our comprehension of those processes has therefore a
large impact.

In this Letter, we report the very first detection of a doubly-deuterated 
isotopomer of methanol (CHD$_2$OH), with 15 detected lines,
towards the low-mass protostar IRAS16293$-$2422.  We also report the
detection of the two singly-deuterated forms (CH$_2$DOH and CH$_3$OD)
of methanol towards the same object.  We compare the derived
fractionation ratios as well as the formaldehyde fractionation
(\cite{Loinard00} 2000) with predictions based on active grain chemistry.


\section{Observations and results}\label{sec-obs}

Using the IRAM 30-meter telescope (Pico Veleta, Spain),
we detected the 23 CH$_2$DOH lines, 6 CH$_3$OD lines, and 15 CHD$_2$OH
lines reported in Table
\ref{table}. The telescope was pointed at the coordinates
$\alpha$(2000) = 16$^{\rm h}$32$^{\rm m}$22.6$^{\rm s}$ and
$\delta$(2000) = -24$^o$28$'$33.0$''$.
The observations were performed in November 2001 and 
May 2002.
Four receivers were used simultaneously at 3, 2, 1.3 and 1.1 mm
with typical system
temperatures of about 100, 200, 400 and 800 K respectively. These
receivers
were connected to
an autocorrelator divided in up to eight units depending on
the setting.
The telescope beam width varies between 30$''$ at 83 GHz and 11$''$
at 226 GHz.
All observations were
performed using the wobbler switching mode with an OFF position 4$'$
from the center.
The pointing accuracy was monitored regularly on strong extragalactic
continuum sources
and found to be better than 3$''$. Our spectra were obtained with
integration times ranging from
70 to 400 min for some CHD$_2$OH lines.

\begin{figure}[!ht]
\includegraphics[width=6.1cm,angle=-90]{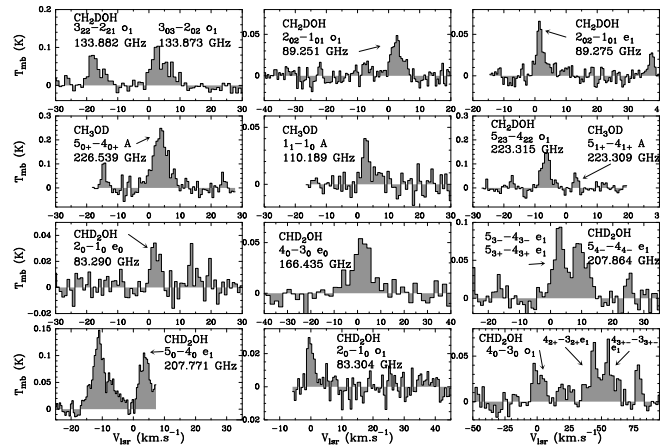}
\caption{Examples of CH$_2$DOH, CH$_3$OD and CHD$_2$OH lines.
The intensities are reported in main-beam brightness temperature.}
\label{lines}
\end{figure}

Examples of observed spectra are shown in Fig. \ref{lines}.
The emission was detected at the v$_{LSR}$ of the source, namely 3.9
km s$^{-1}$.
The lines have FWHM of $\sim 5$ km s$^{-1}$, very similar to the
linewidths observed for the HDCO and D$_2$CO lines in IRAS16293
(\cite{Loinard00} 2000).
The measured intensities, linewidths and main-beam
temperatures are reported in Table \ref{table}.

\begin{table}[!ht]
\caption{Parameters of the observed CH$_2$DOH, CH$_3$OD and CHD$_2$OH
transitions. The
fluxes were derived using gaussian fits, and the uncertainty given is
$\sqrt{{\sigma_{\textrm{stat}}}^2 + {\sigma_{\textrm{cal}}}^2}$ where ${\sigma_{\textrm{stat}}}$
is the statistical error and $\sigma_{\textrm{cal}}$ the calibration
uncertainty (15\%).
The frequencies for CHD$_2$OH are from Mukhopadhyay (in prep.). 
The line strengths (given in Deb$^2$) used to interpret the data are
from J. Pearson (private communication) for CH$_2$DOH, from
\cite{Anderson88} (1988) for CH$_3$OD, and
from \cite{Mukhopadhyay98} (1998) for CHD$_2$OH.
}
\scriptsize
\begin{tabular}{ccccccc}
\hline
\noalign{\smallskip}
Frequency & Transition  &   $\mu^2$S    &  E$_{\rm up}$ & $\int{{\rm
T}_{\rm mb}dv}$ & T$_{\rm mb}$ &  $\Delta$v \\
\tiny GHz &             &        & \tiny cm$^{-1}$ &  \tiny K.km.s$^{-1}$       
&  \tiny  K   & \tiny  km.s$^{-1}$\\
\noalign{\smallskip}
\hline
\noalign{\smallskip}
\tiny  CH$_2$DOH & & & & & & \\
\noalign{\smallskip}
\hline
\noalign{\smallskip}
89.2512  & 2$_{0,2}$-1$_{0,1}$  o$_1$   &    0.7  &   17.2  &
0.19$\pm$0.04  &  0.04 & 4.3$\pm$0.6\\
89.2754  & 2$_{0,2}$-1$_{0,1}$  e$_1$   &    0.7  &   13.8  &
0.16$\pm$0.03  &  0.06 & 2.6$\pm$0.4\\
89.4079  & 2$_{0,2}$-1$_{0,1}$  e$_0$   &    0.8  &    4.5  &
0.31$\pm$0.05  &  0.06 & 4.7$\pm$ 0.2\\
110.1054 &  9$_{1,8}$-9$_{0,9}$  o$_1$   &    3.3  &   83.4  &
0.42$\pm$0.08  &  0.05 & 7.4$\pm$0.9\\
133.8473 &  3$_{0,3}$-2$_{0,2}$  e$_1$   &    1.1  &   18.3  &
0.60$\pm$0.11  &       &           \\
133.8729 &  3$_{0,3}$-2$_{0,2}$  o$_1$   &    1.1  &   21.7  &
0.60$\pm$0.11  &       &           \\
133.8818 &  3$_{2,2}$-2$_{2,1}$  o$_1$   &    0.6  &   33.6  &
0.34$\pm$0.07  &  0.07 & 4.3$\pm$0.9\\
133.8929 &  3$_{2,2}$-2$_{2,1}$  e$_1$   &    0.6  &   27.4  &
0.67$\pm$0.14  &  0.05 &12.2$\pm$2.2\\
133.8974 &  3$_{2,1}$-2$_{2,0}$  o$_1$   &    0.6  &   33.6  &
0.30$\pm$0.07  &  0.08 & 3.4$\pm$0.6\\
133.9302 &  3$_{2,1}$-2$_{2,0}$  e$_1$   &    0.6  &   27.4  &
0.37$\pm$0.15  &  0.07 & 4.8$\pm$2.2\\
134.0655&  3$_{0,3}$-2$_{0,2}$  e$_0$   &    1.2  &    8.9  &
0.67$\pm$0.33  &  0.10 & 6.0$\pm$3.3\\
134.1124&  3$_{2,2}$-2$_{2,1}$  e$_0$   &    0.7  &   20.2  &
0.43$\pm$0.15  &  0.06 & 6.4$\pm$2.3\\
134.1854&  3$_{2,1}$-2$_{2,0}$  e$_0$   &    0.7  &   20.2  &
0.26$\pm$0.06  &  0.06 & 4.1$\pm$0.8\\
207.7808&  2$_{1,2}$-3$_{0,3}$  e$_0$   &    0.3  &   15.9  &
1.01$\pm$0.25  &  0.11 & 8.4$\pm$2.5\\
223.0713&  5$_{2,3}$-4$_{1,4}$  e$_1$   &    0.6  &   33.6  &
1.13$\pm$0.42  &  0.17 & 6.3$\pm$2.5\\
223.1073&  5$_{0,5}$-4$_{0,4}$  o$_1$   &    1.8  &   35.1  &
0.51$\pm$0.18  &  0.15 & 3.1$\pm$1.2\\
223.1283&  5$_{2,4}$-4$_{2,3}$  e$_1$   &    1.4  &   40.8  &
0.54$\pm$0.26  &  0.22 & 2.3$\pm$0.7\\
223.1311&  5$_{4,1}$-4$_{4,0}$  o$_1$   &    0.6  &   79.4  &
0.41$\pm$0.16  &  0.08 & 4.8$\pm$2.1\\
223.1311&  5$_{4,2}$-4$_{4,1}$  o$_1$   &    0.6  &   79.4  &
0.41$\pm$0.16  &  0.08 & 4.8$\pm$2.1\\
223.1537&  5$_{3,3}$-4$_{3,2}$  o$_1$   &    1.0  &   60.8  &
0.58$\pm$0.09  &       &           \\
223.1537&  5$_{3,2}$-4$_{3,1}$  o$_1$   &    1.0  &   60.8  &
0.58$\pm$0.09  &       &           \\
223.3154&  5$_{2,3}$-4$_{2,2}$  e$_1$   &    1.4  &   40.8  &
0.38$\pm$0.09  &  0.16 & 2.2$\pm$0.5\\
223.4223&  5$_{2,4}$-4$_{2,3}$  e$_0$   &    1.8  &   33.6  &
0.55$\pm$0.09  &  0.10 & 5.2$\pm$0.3\\
\noalign{\smallskip}
\hline
\noalign{\smallskip}
\tiny  CH$_3$OD & & & & & & \\
\noalign{\smallskip}
\hline
\noalign{\smallskip}
110.1889 & 1$_{1}$-1$_{0}$     &  1.6   &   7.8   &     0.10$\pm$0.02 & 0.04 &  2.3$\pm$0.5\\
110.2626 & 2$_{1}$-2$_{0}$     &  2.7   &  10.8   &     0.17$\pm$0.04 & 0.03 &  4.5$\pm$1.1\\
110.4758 & 3$_{1}$-3$_{0}$     &  3.8   &  15.4   &     0.30$\pm$0.05 &      &             \\
133.9254 & 1$_{1-}$-1$_{0+}$   &  3.2   &   6.0   &     0.55$\pm$0.18 & 0.06 &  8.4$\pm$2.8\\
223.3086 & 5$_{1+}$-4$_{1+}$   &  3.4   &  26.8   &     0.09$\pm$0.03 & 0.06 &  1.4$\pm$0.4\\
226.5387 & 5$_{0+}$-4$_{0+}$   &  3.5   &  22.7   &     1.32$\pm$0.21 & 0.23 &  5.5$\pm$0.3\\
\noalign{\smallskip}
\hline
\noalign{\smallskip}
\tiny CHD$_2$OH & & & & & & \\
\noalign{\smallskip}
\hline
\noalign{\smallskip}
83.1292   &   2$_{0}$-1$_{ 0}$  e$_1$   &    1.4    &   16.98 &
0.07$\pm$0.02   & 0.02  &  3.1$\pm$0.7\\
83.2895   &   2$_{0}$-1$_{ 0}$  e$_0$   &    1.4    &    4.17 &
0.09$\pm$0.02   & 0.03  &  3.2$\pm$0.6\\
83.3036   &   2$_{0}$-1$_{ 0}$  o$_1$   &    1.4    &   10.33 &
0.07$\pm$0.01   & 0.03  &  2.5$\pm$0.5\\
166.234   &   4$_{0}$-3$_{ 0}$  e$_1$   &    2.8    &   26.69 &
0.39$\pm$0.08   &       &             \\
166.271   &   4$_{2-}$-3$_{2-}$  e$_1$  &    2.1    &   35.65 &
0.22$\pm$0.07   & 0.04  &  5.1$\pm$1.8\\
166.297   &   4$_{3+}$-3$_{3-}$  e$_1$  &    1.2    &   46.54 &
0.10$\pm$0.08   & 0.03  &  3.2$\pm$2.2\\
166.298   &   4$_{3-}$-3$_{3-}$  e$_1$  &    1.2    &   46.54 &
0.10$\pm$0.08   & 0.03  &  3.2$\pm$2.2\\
166.304   &   4$_{2+}$-3$_{2+}$  e$_1$  &    2.1    &   35.65 &
0.22$\pm$0.06   & 0.06  &  3.4$\pm$0.9\\
166.327   &   4$_{0}$-3$_{0}$  o$_1$    &    2.8    &   20.04 &
0.17$\pm$0.10   &       &             \\
166.435   &   4$_{0}$-3$_{0}$  e$_0$    &    2.8    &   13.89 &
0.48$\pm$0.13   & 0.05  &  8.7$\pm$2.8\\
207.771   &   5$_{0}$-4$_{0}$  e$_1$    &    3.5    &   33.63 &
0.44$\pm$0.11   & 0.09  &  4.4$\pm$1.1\\
207.827   &   5$_{2-}$-4$_{2-}$  e$_1$  &    2.9    &   42.59 &
0.19$\pm$0.11   & 0.07  &  2.5$\pm$1.3\\
207.864   &   5$_{4-}$-4$_{4-}$  e$_1$  &    1.3    &   68.41 &
0.43$\pm$0.13   & 0.07  &  5.6$\pm$1.8\\
207.868   &   5$_{3-}$-4$_{3-}$  e$_1$  &    2.2    &   53.48 &
0.17$\pm$0.10   & 0.05  &  3.4$\pm$1.1\\
207.869   &   5$_{3+}$-4$_{3+}$  e$_1$  &    2.2    &   53.48 &
0.17$\pm$0.10   & 0.05  &  3.4$\pm$1.1\\
\noalign{\smallskip}
\hline
\noalign{\smallskip}
\end{tabular}
\label{table}
\end{table}


\section{Derivation of the column densities}

\begin{figure}
\includegraphics[width=9cm]{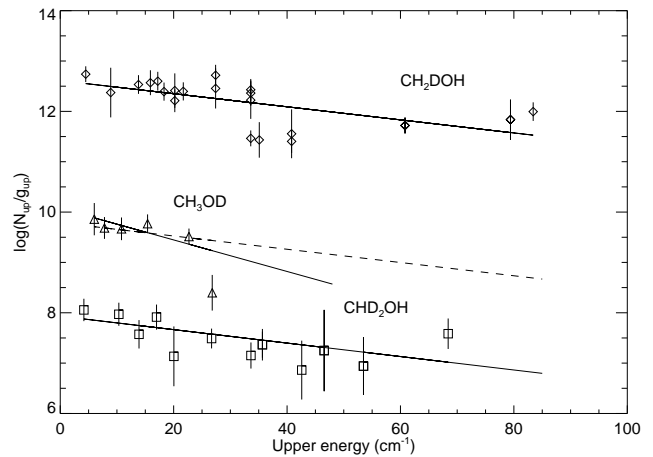}
\caption{Rotational diagrams for the observed transitions of 
CH$_2$DOH (diamonds),
CH$_3$OD
(triangles), and CHD$_2$OH (squares).
For the sake of clarity the CH$_3$OD and CHD$_2$OH points have been
shifted by -2 and -4, respectively. The dashed line is
the result of the fit obtained by fixing the rotational temperature at 47.5 K.}
\label{rotdiag}
\end{figure}

We derived the total column density of the three species from the
rotational diagrams presented in fig. \ref{rotdiag}.
The observed scattering is likely due to non-LTE and opacity effects.
Note that the level column densities are averaged on a
$10''$ beam, i.e. the smallest beam of our observations,
following the suggestion by \cite{vanDishoeck95} (1995) of enhanced methanol
emission in the central $\leq 10''$ region of IRAS16293.
The derived rotational temperatures and overall column densities for the
three molecules are reported in Table
\ref{table2}.
\begin{table}
\caption{$10''$-averaged column density N(Molecule)
and rotational temperature T$_{\rm rot}$
of CH$_2$DOH, CH$_3$OD and CHD$_2$OH respectively.
$^a$Assuming T$_{\rm rot}$=47.5 K (see text).}
\begin{tabular}{ccc}
\hline
\noalign{\smallskip}
Molecule & N(Molecule) (cm$^{-2}$) & T$_{\rm rot}$ (K) \\
\noalign{\smallskip}
\hline
\noalign{\smallskip}
CH$_2$DOH & $(3.0 \pm 0.6) \times 10^{15}$ & $ 48 \pm 3 $ \\
CH$_3$OD  & $(1.5 \pm 0.7) \times 10^{14}$ & $20 \pm 4$ \\
CH$_3$OD$^a$  & $(2.8 \pm 0.3) \times 10^{14}$ & 47.5 \\
CHD$_2$OH & $(6.0 \pm 2.2) \times 10^{14}$ & $ 47 \pm 7$ \\
\noalign{\smallskip}
\hline
\noalign{\smallskip}
\end{tabular}
\label{table2}
\end{table}
The CH$_3$OD rotational temperature is significatively lower
compared with the other two, very probably because the detected lines 
have lower upper level energies.
We therefore estimated the  CH$_3$OD column density assuming 
the same rotational temperature of the two other species,
namely 47.5 K and find the value $(2.8 \pm 0.3) \times
10^{14}$~cm$^{-2}$.
Using the methanol column
density derived by \cite{vanDishoeck95} (1995) ($3.5 \times 10^{15}$
cm$^{-2}$ with
a source size of 10$''$), we finally obtain the following
fractionation ratios:

[CH$_2$DOH]/[CH$_3$OH] = $0.9 \pm 0.3$

[CH$_3$OD]/[CH$_3$OH] = $0.04 \pm 0.02$

[CHD$_2$OH]/[CH$_3$OH] = $0.2 \pm 0.1$

[CH$_2$DOH]/[CH$_3$OD] = $20 \pm 14$

[CHD$_2$OH]/[CH$_2$DOH] = $0.2 \pm 0.1$ .

If the CH$_{3}$OD column density derived with T$_{\rm rot}$=47.5 K  is taken,
the ratios involving this species change by a factor of two,
reflecting the increased abundance of this molecule.

\section{Discussion and conclusions}

Our most dramatic result is the detection in IRAS16293 of a form of
doubly-deuterated methanol along with the detection of both possible
singly-deuterated isotopomers of this molecule.  Up to now, only a
tentative detection of CH$_3$OD has been reported in a low-mass
protostellar source (\cite{vanDishoeck95} 1995).  Singly-deuterated 
methanol has been definitely observed towards Orion
(\cite{Mauersberger88} 1988, where CH$_3$OD/CH$_3$OH $\sim$ 0.01-0.06 
and \cite{Jacq93} 1993, where CH$_{2}$DOH/CH$_3$OH $\sim$ 0.04) and 
SgB2 (\cite{Gottlieb79} 1979, with CH$_3$OD/CH$_3$OH $\sim$ 0.01).  
Equally strikingly, we find the deuterated
forms of methanol to possess a total abundance greater than the main
isotopomer in IRAS16293, even without the contribution of the
doubly-deuterated isotopomer CH$_{2}$DOD!  To date, no other molecule
has been observed to show such extreme deuterium fractionation.

\begin{figure}
\includegraphics[width=9cm]{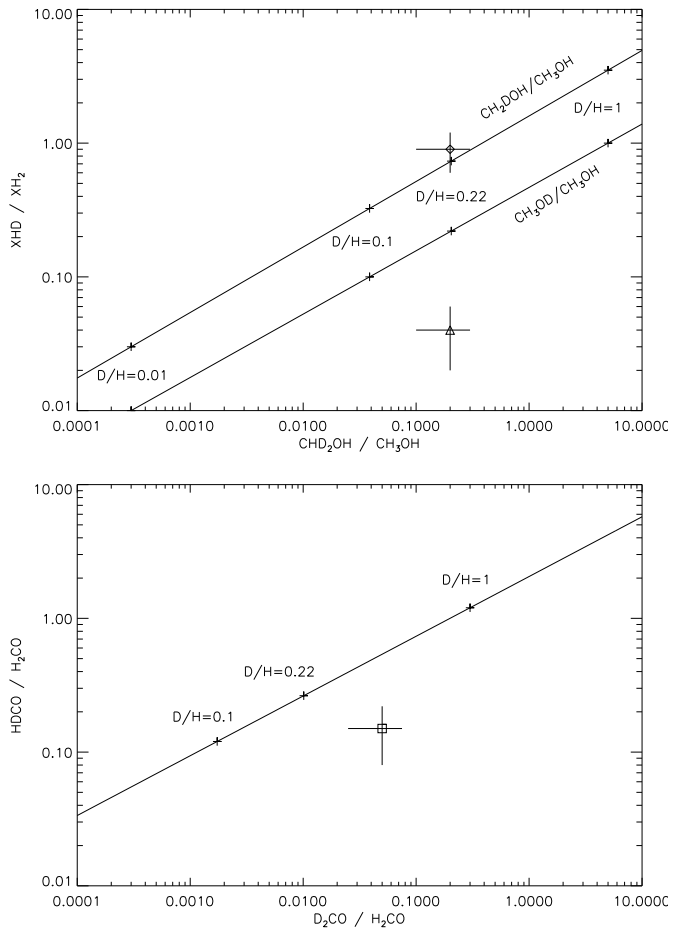}
\caption{Model predictions (solid lines, adapted from CTR97) of 
abundance ratios between
singly-deuterated and doubly-deuterated isotopomers with respect to
normal species are plotted vs one another for methanol and
formaldehyde.  The predictions are obtained as functions of the
gas-phase atomic D/H ratio.  The observed ratios of \cite{Loinard00} (2000)  
are also shown.
Upper panel: CH$_{2}$DOH/CH$_3$OH (diamond) and CH$_3$OD/CH$_3$OH (triangle) versus
CHD$_{2}$OH/CH$_3$OH.
Lower panel : HDCO/H$_2$CO versus D$_2$CO/H$_2$CO. }
\label{ratios}
\end{figure}

As discussed in the Introduction, the abundances of deuterated
methanol and deuterated formaldehyde provide a strong test of models
involving active grain chemistry.  The basic hypothesis behind these
models is that formaldehyde and methanol form by the hydrogenation of
CO accreted onto the grains via reactions with atomic hydrogen
(\cite{Tielens82} 1982; \cite{Charnley97} 1997, hereafter
CTR97).  The enhanced deuteration is caused by an enhanced
(atomic) D/H ratio in the gas during the era of mantle formation
(\cite{Tielens83} 1983).  The hydrogenation and deuteration of CO is
predicted to form H$_2$CO first (CO $\rightarrow$ HCO $\rightarrow$
H$_{2}$CO) and subsequently CH$_3$OH (H$_{2}$CO $\rightarrow$
H$_{3}$CO $\rightarrow$ CH$_{3}$OH). With some simplifying assumptions, this leads directly to
predictions for steady-state ratios of singly- and doubly-deuterated
formaldehyde and methanol to their normal isotopic forms in terms of
the relative accretion rates of H and D with respect to CO as free
parameters (CTR97).  The relative accretion rate of H with respect to
CO can be derived from the observation of the CO/CH$_3$OH and
H$_2$CO/CH$_3$OH abundance ratios.  The predictions for the
fractionation ratios then depend only on the relative accretion rate
of D with respect to CO, or equivalently on the D/H atomic abundance
ratio in the accreting gas. Figure \ref{ratios} shows
fractionation ratios predicted by the CTR97 model.  In particular,
the calculated ratios of singly-deuterated isotopomers to normal
species are plotted against the analogous ratios for doubly-deuterated
species, both as functions of the D/H atomic ratio in the gas.
As seen in the upper panel, the CH$_{2}$DOH and CHD$_{2}$OH observations are
compatible with an atomic D/H ratio of 0.2 in the accreting gas. 
The CH$_3$OD abundance falls short of the predicted value for a D/H ratio
of 0.2.  However, the gas phase abundance of CH$_3$OD in the hot core can
be affected by gas phase ion-molecule reactions.  Specifically,
protonation of methanol by H$_3^+$ or H$_3$O$^+$ followed by dissociative electron
recombination back to methanol will drive the CH$_3$OD/CH$_3$OH ratio to the
deuterium fractionation of molecules in the warm gas, which is very low
(CTR97).  The timescale for this process is some 3$\times$10$^4$ yrs which is
comparable to the lifetime of IRAS16293 ($\sim$ 2$\times$10$^4$, \cite{Ceccarelli00} 2000). 
We note that this chemical reshuffling of the deuterium will not affect
the deuterium fractionation on the methyl group (eg., CH$_2$DOH, CHD$_2$OH). 
The CTR97 model, however, has some difficulties in explaining the observed
formaldehyde fractionation ratios
towards IRAS16293 (\cite{Loinard00} 2000). In particular, the value of
atomic D/H
compatible with CH$_{2}$DOH and CHD$_{2}$OH is reasonably compatible
with HDCO but results in too low a fractionation ratio
for D$_{2}$CO by a factor of 5 or so.
Possible gas-phase alterations have not been considered.
The formaldehyde discrepancy suggests that at least some of the
assumptions in the CTR97 model may be too drastic.
For example, the CTR97 model contains the approximation
that only the accreted H, D and CO are important in the formation of
formaldehyde and methanol on the grain surfaces, and that
no other reactions compete with
formaldehyde and methanol formation.

Very recently, \cite{Caselli02} (2002) [hereafter CSS02] proposed a
somewhat more detailed but related model for the formation of formaldehyde,
methanol, all of their deuterated isotopomers, and selected other
species on grains. In their model, CSS02 consider accretion
onto grains of H, D, CO and O, followed by a comprehensive set of
reactions to form H$_2$O, H$_2$, CO$_2$, H$_2$CO, CH$_3$OH and
all singly- and multiply-deuterated isotopomers of these species.
Their model differs slightly from that 
of CTR97 in that it specifically includes small differences
in the barriers to reaction between non-deuterated and deuterated
species. Yet, in the so-called accretion limit and with the same
assumption of rapid diffusion rates, the CSS02 model
should yield approximately the same results as the CTR97 model
given the same set of chemical reactions and physical conditions.
Predictions for methanol and formaldehyde fractionation ratios are
indeed similar to the CTR97 model and substantially the same
discrepancies remain.  In particular, for D/H = 0.3, a temperature of
10 K, and so-called high-density conditions, CSS02 agree approximately
with our observed fractionation ratios for CH$_{2}$DOH and 
CHD$_{2}$OH, but produce
approximately 5 times too much CH$_{3}$OD, a value similar to that of
CTR97 with D/H = 0.2 and no subsequent gas-phase chemistry.  For
formaldehyde, CSS02 obtain a fractionation ratio for HDCO that is
twice the observed value and a fractionation ratio for
D$_{2}$CO that is a factor of two below the observed value.

In summary, neither model is in good agreement with all of our data.
More importantly perhaps, the CTR97 and CSS02 models
require a D/H atomic ratio in the range 0.2-0.3,
which is a significantly larger value than can be produced by current
gas-phase models, even in the presence of a large CO depletion
(e.g. \cite{Roberts00a} 2000a).
Future progress will probably require more detailed chemical models in
which  gas-phase and surface chemistry occur simultaneously.
Although such models are currently in existence, they do not yet contain
fractionation processes.

\begin{acknowledgements}
We wish to thank John Pearson for providing us the CH$_2$DOH line
strengths and Paola Caselli
for the many fruitful discussions on her model. E. Herbst
acknowledges the support of the National Science
Foundation for his research program in astrochemistry.

\end{acknowledgements}

{}


\end{document}